\newif\ifcameraready
\camerareadytrue

\ifcameraready
    \documentclass[sigconf,screen]{acmart}
\else
    \documentclass[sigconf,review,anonymous]{acmart}
\fi

\copyrightyear{2026}
\acmYear{2026}
\setcopyright{cc}
\setcctype{by}
\acmConference[ASE '26]{Proceedings of the 41st IEEE/ACM International Conference on Automated Software Engineering}{October 12--16, 2026}{Munich, Germany}
\acmBooktitle{Proceedings of the 41st IEEE/ACM International Conference on Automated Software Engineering (ASE '26), October 12--16, 2026, Munich, Germany}
\acmDOI{10.1145/3832783.3834352}
\acmISBN{979-8-4007-2882-2/2026/10}
\acmSubmissionID{ase26main-p379-p}
\received{2026-03-25}
\received[accepted]{2026-06-18}

\usepackage[utf8]{inputenc}
\usepackage{cleveref}
\usepackage{color, colortbl}
\usepackage{listings}
\usepackage{multirow}
\usepackage{siunitx}
\usepackage{tcolorbox}
\usepackage[textsize=tiny]{todonotes}
\usepackage{xspace}
\usepackage{enumitem}
\newcolumntype{H}{>{\setbox0=\hbox\bgroup}c<{\egroup}@{}}
\usepackage{siunitx}

\makeatletter
\newcommand\notsotiny{\@setfontsize\notsotiny\@vipt\@viipt}
\makeatother

\crefname{lstlisting}{listing}{listings}
\Crefname{lstlisting}{Listing}{Listings}

\newcolumntype{P}[1]{>{\centering\arraybackslash}p{#1}}

\tcbuselibrary{breakable}

\hypersetup{
    hidelinks,
    breaklinks=true,
}

\setlist[description,enumerate]{
    labelsep=0.2em,
    leftmargin=\parindent,
}

\newcommand{\paragraphbf}[1]{\smallskip \noindent \textbf{#1.}}

\newcommand{\infobox}[1]{
    \begin{tcolorbox}[
        breakable,
        pad at break*=2mm,
        colback=gray!10,
        colframe=black,
        notitle,
        boxrule=0.5pt,
        left=2pt,
        right=2pt,
        top=2pt,
        bottom=2pt
    ]
        #1
    \end{tcolorbox}
}

\newcommand\important[1]{{\textbf{\textcolor{red!70!black}{#1}}}}

\newcommand\PhpCve{2021 attack\xspace}
\newcommand\ProftpdCve{CVE-2010-20103\xspace}
\newcommand\VsftpdCve{CVE-2011-2523\xspace}

\newcommand\Rosa{\textsc{Rosa}\xspace}
\newcommand\Rosarum{\textsc{Rosarum}\xspace}

\newcommand\Lily{\textsc{Lily}\xspace}

\newcommand\LilySelective{\textsc{LilySelective}\xspace}

\newcommand\LilyRepoURL{\url{https://github.com/binsec/rosa/tree/lily}\xspace}
\newcommand\LilySWHID{swh:1:rev:2bd97b1c06c315a986d969fddec6049b1cc27118}
\newcommand\ReplicationPackageURL{\url{https://zenodo.org/records/19337349}\xspace}

\newcommand\TotalTargets{13\xspace}
 
\title[Not In My Git Yard: Catching Backdoors at Commit and Release Time]{Not In My Git Yard:\\ Catching Backdoors at Commit and Release Time}

\author{Dimitri Kokkonis}
\correspondingauthor
\email{dimitri.kokkonis@cea.fr}
\orcid{0009-0009-5171-2992}
\affiliation{\institution{Université Paris-Saclay}
    \institution{CEA, List}
    \city{Paris-Saclay}
    \country{France}
}

\author{Michaël Marcozzi}
\email{michael.marcozzi@cea.fr}
\orcid{0000-0002-8087-0537}
\affiliation{\institution{Université Paris-Saclay}
    \institution{CEA, List}
    \city{Paris-Saclay}
    \country{France}
}

\author{Stefano Zacchiroli}
\email{stefano.zacchiroli@telecom-paris.fr}
\orcid{0000-0002-4576-136X}
\affiliation{\institution{LTCI, Télécom Paris}
    \institution{Institut Polytechnique de Paris}
    \city{Palaiseau}
    \country{France}
}

\begin{abstract}

Code-level backdoors---stealthy code changes that grant hidden privileges via secret triggers---pose a persistent threat to open-source software.
Known attempts to inject such backdoors into widely used projects through malicious commits,
tampered release packages,
or compromised third-party dependencies, were stopped only by luck and manual review.
Existing Continuous Integration (CI) pipelines cannot detect these attacks,
and downstream binary analysis tools require substantial manual effort.
In this work, we present \Lily, an automated approach that strengthens open-source development and release processes against backdoor injection. \Lily integrates a backdoor detection mechanism into
(1) CI pipelines to block malicious commits, and
(2) release vetting workflows to prevent tampered releases or compromised dependencies from entering large ecosystems, such as Linux distributions.
\Lily offers two key contributions. First, it enhances CI-compatible fuzzing with the capability to detect triggers of suspicious behavior based on historical and current software executions. This enables fast, precise backdoor detection suitable for both CI and update validation workflows.
Second,
it combines code change analysis with fuzzing data
to precisely point maintainers to backdoor-revealing code regions,
even when release updates modify millions of lines of code.
We also outline five strategies attackers could use to evade \Lily,
and evaluate corresponding defenses.
Our experiments across hundreds of benign and backdoored commits and releases show that \Lily achieves high detection accuracy with low false alarm rates,
reliably identifies malicious code,
resists adversarial attempts,
and would have prevented real-world backdoor incidents.

\end{abstract}
 


\begin{document}

\maketitle

\section{Introduction}
\label{sec:introduction}

\paragraphbf{Context} \emph{Code-level backdoors}~\cite{thomas-2018-backdoors,kokkonis-2025-rosa}---hidden functionalities embedded within source code that allow individuals aware of their presence to obtain elevated privileges or unauthorized features via secret triggers---remain a persistent threat to open-source software. Incidents involving malicious commits to the PHP public repository~\cite{php-backdoor}, tampering with ProFTPD and vsFTPd release packages~\cite{proftpd-backdoor-cve,vsftpd-backdoor-cve}, or the injection of backdoors through a compromised dependency to XZ Utils~\cite{xz-backdoor-cve} have gone undetected for days, being uncovered only through a combination of manual inspection and chance. The consequences of missing such threats are severe, especially as the ever-expanding ecosystem of software dependencies magnifies the potential impact of a single compromise~\cite{decan-2019-dep-networks}.

\paragraphbf{Problem} The current open-source software development and release stack lacks systematic defenses against backdoor injection attacks. Upstream, \emph{Continuous Integration (CI) pipelines} run automated tests---typically on every commit---to ensure code quality through compilation checks and regression testing. \emph{Fuzzing}, already integrated into CI pipelines of major projects~\cite{libsndfile, php, openssl, sudo}, targets crashes and memory safety bugs through brute-force testing of the revised code. Because CI must provide rapid feedback, fuzzing runs are brief, often limited to about 10 minutes~\cite{huang-2024, klooster-2023, fowler-2024}. Existing CI mechanisms, however, cannot detect backdoored commits such as hard-coded credentials or hidden reverse shells. A recent machine-learning approach that analyzes Git metadata~\cite{ganz-2023} could offer a partial solution but still yields many false alarms.
Downstream, state-of-the-art backdoor detection tools~\cite{schuster-2013, shoshitaishvili-2015, thomas-2017-stringer, thomas-2017-humidify, kokkonis-2025-rosa} help end users vet binaries, but demand manual effort to reverse-engineer code or to dismiss false alarms, limiting scalability.

\paragraphbf{Goal and Challenges}
We aim to harden open-source development and release pipelines
by introducing a mechanism that blocks most backdoor injection attacks,
integrating it in two stages:
(1) in CI pipelines,
and (2) in release vetting before updated packages enter major ecosystems such as Linux distributions.
Doing so requires addressing three challenges:
\begin{enumerate}
\item CI and release vetting pipelines operate under strict time constraints, so backdoor detection must be fully automated and able to achieve a high detection rate within these limits to warrant integration. At the same time, any false alarm can halt the development or release process, making high precision essential to maintain developer productivity and preserve maintainer trust.
    \item The code changes to inspect (commit or release diffs) can span thousands to millions of lines, requiring precise reports that pinpoint suspicious locations for CI operators or distribution maintainers.
    \item Broad adoption may prompt adversaries to develop evasion techniques, which must be evaluated and countered.
\end{enumerate}
 
\paragraphbf{Proposal} We propose \Lily, a new approach for preventing backdoor injection in open-source development. \Lily augments existing 10-minute CI fuzzing campaigns with a runtime monitoring mechanism for detecting backdoor triggers, and extends these campaigns to also vet new package releases.
\begin{enumerate}
\item While the revised code is undergoing fuzz testing, \Lily monitors its runtime behavior. It flags a potential backdoor only when both of the following conditions are satisfied: (i) it observes a behavior that was absent from the pre-revision code, and (ii) the observed behavior deviates significantly from the revised code's typical behavior, as characterized by executing the revised code on a regression test suite derived from the fuzzing campaigns conducted when vetting prior revisions.
    \item \Lily correlates static code change information from commits and releases with dynamic data from fuzzing and monitoring,
        enabling precise localization of backdoor-revealing code regions.
    \item We systematically evaluate five possible evasion strategies---CI bypass, fuzzer obstruction,
        segmented or adversarial backdoor patterns and regression test suite poisoning---and assess their feasibility and countermeasures.
\end{enumerate}

\paragraphbf{Evaluation} We evaluate \Lily through 20 ten-minute detection trials on 432 safe and 13 backdoored commits, and on 50 safe and 50 backdoored releases from 13 popular open-source projects. The backdoors come from the \Rosarum corpus---real and synthetic---introduced at ICSE'25 as a benchmark for detection tools, following established fuzzing evaluation practices~\cite{kokkonis-2025-rosa, schloegel-2024}.
    \Lily achieves an average detection rate of \num{90}\% at the commit level and \num{83}\% at the release level, indicating that it likely would have prevented three real backdoor injection incidents (\ProftpdCve, \VsftpdCve, and the PHP attack~\cite{php-backdoor}). False alarm rates remain low---0.2\% for safe commits and 4.3\% for safe releases on average---and they occur primarily in predictable scenarios such as large merge commits or test harness modifications.
Our ablation study confirms that reporting a behavior as a potential backdoor only when it is both novel and atypical dramatically reduces false alarms.
Even in worst-case scenarios involving releases differing by thousands to millions of lines,
\Lily consistently reduced manual review to under a dozen lines,
always pinpointing backdoor-revealing code.
Our adversarial analysis identifies possible evasion strategies,
but they require substantial effort,
offer no guaranteed success,
and increase attacker exposure.
Countermeasures, including our hardened mode, improve resilience and prevent regression test suite poisoning.

\paragraphbf{Contributions} To sum up, our main contributions are:
\begin{enumerate}
    \item \Lily, an automated backdoor detection approach
        which systematically hardens open-source development and release pipelines against injection attacks.
        During revised code fuzzing, Lily tracks runtime behavior and flags potential backdoors whenever new execution patterns emerge (with regards to the original code) and fall outside the revised code's normal behavior.
        It further combines code change analysis with fuzzing data
        to precisely guide maintainers to backdoor-revealing code regions,
        even in releases modifying millions of lines of code.
\item A comprehensive evaluation over hundreds of safe and backdoored commits and releases,
        showing high detection rates with low false alarms,
        accurate localization of backdoor-revealing code,
        and robustness under adversarial conditions,
        including several different attack vectors and countermeasures to them.
        It also demonstrates that \Lily would have prevented several real-world backdoor incidents~\cite{proftpd-backdoor-cve,vsftpd-backdoor-cve,php-backdoor}.
\end{enumerate}
 \section{Background}
\label{sec:background}

\subsection{Release Cycle and Continuous Integration}
\label{subsec:background-ci-cd}
In modern software engineering,
project development is commonly organized into \emph{release cycles},
during which \emph{releases}
(i.e., new versions of a program)
are periodically delivered to end users.
Each release provides a set of updates
(e.g., new features, bug fixes).
Between releases,
it is standard practice to rely on a Version Control System (VCS) such as Git~\cite{git}
(whose history is typically public in the case of open-source projects)
to facilitate collaboration and ensure traceability.
This workflow is frequently complemented by a \emph{Continuous Integration} (CI) system,
often called a ``CI pipeline.''
Its primary goal is to provide continuous quality assurance,
by automatically executing a sequence of \emph{jobs} (i.e., tasks)
which check whether recent changes \emph{committed} to the VCS
maintain the integrity of the software.
Typical examples include compiling the software
and running an associated regression test suite.
When a contributor introduces a defect detectable by CI jobs,
the CI pipeline reports an error,
and typically blocks the integration of the faulty change into the upstream codebase.
The individuals responsible for the change are notified,
and are expected to inspect the CI logs,
diagnose the issue,
and submit a correction.

\subsection{Fuzzing in Open-Source Development}
\label{subsec:background-ci-fuzzing}

\emph{Graybox fuzzing}~\cite{godefroid-2020} is an effective automated testing technique that generates test inputs for the Program Under Test (PUT) and uses coverage feedback to guide program exploration~\cite{fioraldi-2020}. Fuzzers monitor the execution of the PUT---often using sanitizers~\cite{song-2019}---and report crashing inputs that reveal issues such as memory errors or assertion failures.
Google's OSS-Fuzz~\cite{oss-fuzz}, launched in 2016, provides free fuzz testing for open source projects, supporting engines such as the AFL++ fuzzer~\cite{fioraldi-2020}. Developers supply a \emph{fuzzing harness} linking the fuzzer to the project API; for example, libpng's harness~\cite{libpng} loads and processes fuzz-generated PNG files. Harness design strongly affects fuzzing quality, as it determines which parts of the codebase are exercised and under what conditions. As of August 2023, OSS-Fuzz had helped fix over 10{,}000 vulnerabilities in 1{,}000 projects.
 To catch issues as early as possible, some projects also integrate fuzzing in CI pipelines. OSS-Fuzz offers CIFuzz,\footnote{\notsotiny\url{https://google.github.io/oss-fuzz/getting-started/continuous-integration}} which builds each commit, runs a brief fuzzing campaign (typically 10 minutes~\cite{huang-2024, klooster-2023, fowler-2024}), and records any crashes with their stack traces and inputs.
 
\subsection{Injecting Code-Level Backdoors}
\label{subsec:backdoor-detection}

\subsubsection{Code-Level Backdoors}
\label{subsubsec:code-level backdoors}
While backdoors of many forms have been identified in various components of computer systems~\cite{lokhande-2019,fang-2021,kostyuk-2022,chen-2017}, this work concentrates on code-level backdoors~\cite{thomas-2018-backdoors,kokkonis-2025-rosa}. These are hidden functionalities embedded directly within the source code of conventional programs, enabling individuals aware of their existence to gain elevated privileges or unauthorized access by triggering them with a secret input. A notable example is the official vsFTPd distribution, which was once reported~\cite{vsftpd-backdoor-cve} to include a backdoor. In this case, using the string \lstinline{":)"} as the FTP username would open a remote root shell on the compromised system hosting the backdoored vsFTPd.

\subsubsection{Injecting Backdoors in Open-Source Projects}
\label{subsubsec:real-world-attacks}

Although relatively uncommon, code-level backdoor injections in open-source projects have been reported consistently over the years, indicating persistent and organized attempts to compromise software that often underpins critical infrastructure. Among the most notable incidents, attackers infiltrated the XZ Utils library in 2024~\cite{lins-2025-xz-attack-path} (CVE-2024-3094), embedding a multi-stage backdoor into its Git repository. This compromise weakened OpenSSH authentication (as OpenSSH depends on XZ Utils), enabling unauthorized SSH access. The attack was ultimately detected due to performance anomalies. In 2021, the {PHP} Git server was breached~\cite{php-backdoor}, and malicious commits introduced a backdoor into the HTTP server bundled with the PHP interpreter. This vulnerability allowed remote code execution via crafted HTTP headers, but it was detected before reaching an official release, so no CVE was issued. Earlier, between 2010 and 2011, attackers compromised the {vsFTPd} and {ProFTPD} distribution servers~\cite{vsftpd-backdoor-cve, proftpd-backdoor-cve} (CVE-2011-2523 and CVE-2010-20103), distributing releases with hard-coded backdoors for several days.

From such incidents, three primary backdoor injection attack scenarios on open-source projects emerge:

\begin{description}
    \item[Injection via malicious commits:]
        achieved by getting harmful commits accepted into the upstream repository, or by pushing them from compromised maintainer accounts. This leaves traces in version control history and immediately compromises the software, as seen in the PHP incident.

    \item[Out-of-repository injection:]
        achieved by infecting official release packages
        with the backdoor, as in the vsFTPd and ProFTPD cases.

    \item[Software supply-chain injection:]
         a backdoor inserted into a compromised dependency propagates to downstream components, exploiting the complexity of modern dependency networks. The XZ Utils incident illustrates this: embedded in thousands of projects, its backdoor activated only when used by OpenSSH.
\end{description}

 \section{The \Lily Approach}
\label{sec:approach}
\subsection{General Overview} We introduce \textbf{\Lily},
a novel approach to preventing code-level backdoor injections in open-source projects.
At its core,
\Lily employs a code analysis based on graybox fuzzing which compares two versions of the same PUT.
By examining the differences introduced in the newer version,
\Lily determines whether these changes may include a backdoor.

\Lily can be deployed at different stages of open-source development workflows (see discussion in \Cref{subsec:deployment}). It consists of three components implementing a three-step approach, which are summarized below and detailed in \Cref{subsec:representative-input-corpus,subsec:lily-oracle,subsec:suspicious-commit-report}. \textbf{\Cref{subsec:example} illustrates the approach on a concrete example}.

\begin{description}
    \item[1. Standard behavior identifier:]
        establishes a baseline of normal behaviors for the \emph{new version} of the PUT by executing it on a regression test suite derived from fuzzing campaigns for earlier PUT revisions. Inspired by multi-version execution~\cite{hosek-2015-varan} and \Rosa~\cite{kokkonis-2025-rosa} (see \Cref{sec:related-work}), \Lily characterizes behavior using system-call profiles, i.e., the sets of system-call types observed during each test execution.\footnote{Because system calls mediate all external interactions, any meaningful code-level backdoor must invoke them and can therefore be detected through monitoring (see \Cref{subsec:attacker-multi-step-backdoor,subsec:attacker-common-syscalls} for potential evasions).}

    \item[2. Backdoor detection oracle:] during a fuzzing campaign on the new version of the PUT, this oracle reports a backdoor only if both conditions hold: (i) a fuzzer-generated input triggers a system call profile that differs from the normal profiles identified by the standard behavior identifier, and (ii) the same input does not trigger the corresponding behavior when executed on the previous version of the PUT. Requiring both conditions serves to reduce false alarms by filtering out legitimate new behaviors introduced by the revision (condition 1) as well as atypical behaviors already present in the previous version (condition 2).

    \item[3. Suspicious code tracer:]
        when the oracle flags a potential backdoor, the tracer highlights the modified source lines that are most likely associated with the suspicious behavior, by filtering out benign changes to produce cleaner, more focused logs. It correlates the non-standard behaviors observed in the new version of the PUT with the source-code differences between the two versions.
\end{description}

\subsection{Deploying \Lily During Development} \label{subsec:deployment} \Lily can be deployed at either of two (or both) critical stages of the open-source lifecycle
to mitigate the three previously described injection scenarios (\Cref{subsubsec:real-world-attacks}):

\begin{description}
    \item[During development:]
        \Lily is run at the CI stage,
        complementing existing fuzzing jobs (as in OSS-Fuzz's~\cite{oss-fuzz} CIFuzz).
        By vetting the code changes introduced by each new commit, it provides the capability to block backdoor-injecting commits.

    \item[After release updates:]
        \Lily is run when end users integrate new versions of an upstream project,
        to detect supply-chain and out-of-repository injections.
        \emph{In our evaluation (\Cref{sec:evaluation}),
        we consider the scenario in which maintainers of major Linux distributions
        (Debian and Ubuntu)
        automatically vet new package releases for inclusion in an upcoming distribution version,
        comparing them against the old package versions they previously shipped
        (including the same older dependency versions).
        Unlike vetting individual commits,
        reviewing a package release means assessing many bundled changes at once.
        Still, because distribution maintainers must handle large numbers of external packages,
        similar CI-level constraints apply:
        short fuzzing time, no manual setup, low false alarms.}
\end{description}

\Lily is designed to prevent backdoor injections as early as possible in active projects. As such, \textbf{when deployed on an existing open-source project, it will only detect backdoors introduced in future commits or releases, not those already present in the codebase}. Detecting preexisting backdoors typically requires a one-time, clean-slate audit involving more extensive analysis, for which state-of-the-art tools such as \Rosa~\cite{kokkonis-2025-rosa} are better suited. In practice, a project can first be audited with \Rosa and then protected by \Lily against future backdoor injections.

\subsection{Identifying Standard Behaviors}
\label{subsec:representative-input-corpus}

Graybox fuzzers generate test inputs through mutation: new inputs are produced by randomly modifying existing ones. A feedback loop guides this process. As mutated inputs execute on the PUT, only those that increase control-flow edge coverage are retained as seeds. These seeds then serve as the basis for further mutations, enabling the fuzzer to progressively explore deeper behaviors of the PUT. A campaign begins with a user-provided seed corpus, proceeds through many iterations of mutation and execution, and ultimately outputs the retained seeds.

Integrating \Lily into an existing development or release pipeline begins with running an initial graybox fuzzing campaign on the latest commit or release. The resulting seeds are saved as the first regression suite to be consumed by the standard behavior identifier. Starting from this baseline, \Lily fuzzes each subsequent commit or release as it becomes available to detect potential injection attempts. For every new version, it reuses all seeds retained during the previous campaign---both as the initial seed corpus for fuzzing and as the regression suite for the standard behavior identifier. This allows the corpus and test suite to evolve organically with the codebase through the fuzzer's feedback loop.

For a given commit or release under analysis, the standard behavior identifier executes the regression suite on the revised PUT and collects all distinct system call profiles observed during execution. A system call profile is the set of system call types triggered during the execution of an input, among those provided by the operating system API (e.g., \texttt{read}, \texttt{kill}, \texttt{open} in Linux). The resulting set of profiles provides a compact yet expressive characterization of the revised PUT's standard behaviors.

\subsection{Vetting Changes for Backdoor Injections}
\label{subsec:lily-oracle}
When vetting a specific commit or release, the newer version of the PUT is fuzzed with a graybox fuzzer for a duration consistent with the target use case and practical constraints. In our evaluation, \Lily uses a 10-minute fuzzing window with AFL++, which aligns with state-of-the-art CI fuzzing practices such as those used in CIFuzz. Longer fuzzing durations can be configured when CI runners allow it or when end users have sufficient computational resources to vet release updates more extensively.

Each input seed produced by the fuzzer during this vetting phase is analyzed in two steps.
First, the seed is traced on the newer PUT, and its resulting system call profile is obtained. This profile is then compared to the set of standard profiles identified by the standard behavior identifier (\Cref{subsec:representative-input-corpus}) for the newer version of the PUT. If an exact match is found, the input is labeled \emph{safe}.
Second, if the profile does not match any standard behavior on the newer PUT, the input is traced again on the older version. If the same non-standard profile also appears there, the behavior is not caused by the vetted change, and the input is likewise labeled \emph{safe}. Otherwise, the input is labeled \emph{suspicious}, since the vetted change is responsible for introducing this new and atypical behavior---a characteristic frequently associated with backdoor injections~\cite{proftpd-backdoor-cve, vsftpd-backdoor-cve, php-backdoor}. Inputs classified as suspicious are forwarded to the suspicious code tracer to identify the code modifications responsible for the anomalous behavior.

\textbf{A key aspect of this design} is that the standard behavior identifier collects standard system call profiles \textbf{on the {new} version of the PUT}. This \textbf{helps mitigate false positives caused by benign changes} that legitimately add or remove system calls: the standard behaviors are always computed with these modifications already taken into account.

\subsection{Producing Precise Backdoor Reports}
\label{subsec:suspicious-commit-report}
When fuzzing the newer PUT reveals an input that triggers a non-standard system call profile,
and this profile is not reproduced by executing the older PUT with the same input,
\Lily reports a backdoor injection attempt.
To transform these findings into a more interpretable evaluation of suspicious code changes for inclusion in CI logs or release vetting reports,
the suspicious code tracer proceeds as follows:
\begin{enumerate}
    \item \textbf{Isolate the backdoor trigger zone} by recording line coverage for the suspicious input within the modified code segments.
    \item \textbf{Isolate the backdoor impact zone} by using catchpoints for the suspicious system calls and backtraces to reveal the lines of code which lead to them (using GDB~\cite{gdb}).
    \item \textbf{Generate the final backdoor report}, which includes the trigger and impact zones (or their intersection, if present), along with a proof of concept (PoC): namely, the identified triggering input and the resulting non-standard system call profile.
\end{enumerate}

\begin{lstlisting}[
    float,
    caption={
        Example of a \Lily finding report for the vsFTPd backdoor. The lines responsible for the
        suspicious system calls are marked in bold, and the C Library functions emitting these
        system calls are marked in \important{red}.
    },
    captionpos=b,
    label=lst:vsftpd-report,
    basicstyle=\footnotesize\ttfamily,
    frame=single,
    keywordstyle=\bfseries,
    moredelim={[is][keywordstyle]{!!}{!!}},
    escapeinside={<@}{@>},
]
--- a/sysdeputil.c
+++ b/sysdeputil.c
@@ -845,0 +847,23 @@
+int
+vsf_sysutil_extra(void)
+{
+  int fd, rfd;
+  struct sockaddr_in sa;
!!+  if((fd = <@\important{socket}@>(AF_INET, SOCK_STREAM, 0)) < 0)!!
+  exit(1);
+  memset(&sa, 0, sizeof(sa));
+  sa.sin_family = AF_INET;
+  sa.sin_port = htons(6200);
+  sa.sin_addr.s_addr = INADDR_ANY;
!!+  if((<@\important{bind}@>(fd,(struct sockaddr *)&sa,!!
+  sizeof(struct sockaddr))) < 0) exit(1);
!!+  if((<@\important{listen}@>(fd, 100)) == -1) exit(1);!!
+  for(;;)
+  {
!!+    rfd = <@\important{accept}@>(fd, 0, 0);!!
!!+    <@\important{close}@>(0); <@\important{close}@>(1); <@\important{close}@>(2);!!
!!+    <@\important{dup2}@>(rfd, 0); <@\important{dup2}@>(rfd, 1); <@\important{dup2}@>(rfd, 2);!!
!!+    <@\important{execl}@>("/bin/sh","sh",(char *)0);!!
+  }
+}
+
\end{lstlisting}

\subsection{Illustrative Example: The vsFTPd Backdoor}
\label{subsec:example}

To illustrate how \Lily operates on a real-world example, consider the
vsFTPd backdoor~\cite{vsftpd-backdoor-cve} discussed earlier (\Cref{subsubsec:code-level backdoors}).

When fuzzing the infected release of the program to determine whether any
backdoor was introduced compared to the previous official release, the
AFL++ fuzzer~\cite{fioraldi-2020} eventually generates an input that triggers
the backdoor---an FTP username containing the string \lstinline{":)"}. Because
this input activates the backdoor code, it exercises previously unseen
control-flow edges; AFL++ therefore saves it as a seed and passes it to
\Lily for analysis.
This seed produces a combination of system call types (shown in
\Cref{lst:vsftpd-report}) that never occur together when running the same
infected release on all inputs from the existing regression suite---that is,
the seeds AFL++ had accumulated while validating the earlier, non-infected
release of vsFTPd. (Naturally, the fuzzer is unlikely to have saved the
backdoor-triggering input among those earlier seeds, since the backdoor code
did not exist in the previous release.)

Furthermore, the earlier, non-infected version of vsFTPd does not produce the
same system call profile when executed on this input.
\Lily's backdoor detection oracle therefore concludes that:
\begin{enumerate}[label=(\roman*)]
  \item the input that triggers the backdoor induces a non-standard behavior in the infected release, and
  \item this behavior cannot be reproduced on the previous release.
\end{enumerate}
Consequently, a backdoor is reported and the suspicious code tracer:
\begin{enumerate}
  \item collects the source line coverage for the suspicious input,
  \item intersects this coverage with the code differences between the two releases, and
  \item identifies the callsites responsible for the anomalous system calls, highlighting them within the code intersection from step~(2).
\end{enumerate}
As a result, the potentially suspicious modifications are narrowed down to a
single file and seven lines, as shown in \Cref{lst:vsftpd-report}. In contrast,
the raw diff between the previous and infected releases spans five files and
1{,}407 lines of code. In the report, lines emitting suspicious system calls are
displayed in \textbf{bold}, and the corresponding C library functions appear in
\important{red}. This enables the human reviewer to quickly determine that the backdoor exposes a remote shell over a TCP socket, allowing an attacker to connect to the server and execute arbitrary commands. Based on this evidence, the reviewer can confidently confirm the presence of the backdoor, reject the entire release, and discard any artifacts derived from it that could otherwise be reused by \Lily in subsequent iterations (e.g., seeds used to construct future regression test suites).
 \section{Mitigating Adversarial Attacks}
\label{sec:attacker-models}

As with existing CI fuzzing approaches,
\Lily adopts a best-effort strategy.
In practice,
it can detect and block the injection of real vulnerabilities.
However,
there are no hard guarantees that the fuzzer will trigger and identify all injected vulnerabilities.
As with most security defenses,
the primary objective is to \emph{increase the cost of attacks} rather than make them entirely impossible.
Nevertheless, if \Lily is widely adopted,
adversarial attackers may attempt to circumvent its mechanisms.
In this section, we systematically review different attack scenarios,
analyze their potential effectiveness,
and discuss possible countermeasures.

\subsection{Bypassing CI Checks}
\label{subsec:attacker-skip-ci}
In many projects, core maintainers can bypass CI checks by adding a special command in the commit message or by using an option when pushing the commit to the repository. An attacker could exploit this to skip any \Lily runs scheduled in the CI pipeline.
To prevent this, \Lily could run as an independent service provided by the Git hosting platform,
ensuring that backdoor detection happens outside the control of the project's development team,
and notifying all stakeholders should a backdoor be detected
(e.g., by displaying a banner on the repository).
This approach guarantees unbiased detection even if core maintainers are compromised~\cite{lins-2025-xz-attack-path}.

\subsection{Avoiding Backdoor Activation While Fuzzing}
\label{subsec:attacker-fuzzer}
An attacker may attempt to conceal a backdoor in code that is difficult for a fuzzer to reach. To assess this risk, we measured reachable line coverage for 337 code-affecting commits across 12 open-source projects (see \Cref{subsec:benchmark}). Average final coverage is approximately 32\%, although coverage varies substantially across commits: some are fully covered, whereas others receive only limited coverage. Our manual investigation revealed that low coverage primarily stems from limitations of the available developer-provided fuzzing harnesses. In several cases, the affected code resides in components that were not prioritized for fuzz testing, while in others it belongs to legacy functionality for which harnesses were still immature at the time.
Although coverage is inherently constrained by available fuzzing resources, continued advances in computing performance and fuzzing techniques are likely to reduce these limitations over time. Increasing fuzzing budgets for randomly sampled \Lily runs could further raise the cost of evasion trials. Regarding harnesses, we observed a consistent improvement in both quality and scope throughout the histories of the 12 projects we studied; when evaluated with modern harnesses, code-affecting commits typically achieve near-complete coverage. OSS-Fuzz maintainers indeed routinely refine existing harnesses, and recent advances in (semi-)automatic harness generation~\cite{chen-2023-hopper, sherman-2025-harness} may further strengthen these efforts.

More advanced attacks may use anti-fuzzing techniques to prevent backdoor activation~\cite{jung-2019-anti-fuzzing,sarafov2025tephra}, such as cryptographic checks like those in the XZ Utils incident~\cite{lins-2025-xz-attack-path}. However, CI and continuous fuzzing systems (e.g., OSS-Fuzz~\cite{oss-fuzz}) routinely generate coverage reports to help developers improve fuzzing harnesses and increase PUT coverage. Deliberately making parts of the PUT hard to reach could therefore backfire by drawing attention to newly uncovered code containing a backdoor. This may help explain why the attackers disabled CI fuzzing in the XZ Utils case.

\subsection{Injecting the Backdoor in Multiple Steps}
\label{subsec:attacker-multi-step-backdoor}

\paragraph{The PHP Attack.}
In the PHP incident~\cite{php-backdoor},
attackers tried to hide the backdoor by spreading it across multiple commits.
The first commit (1) injected the backdoor,
the second (2) reverted it (appearing to ``fix'' the issue),
and a third (3) reverted the revert,
restoring the original backdoor.
We reproduced this scenario and found that it had no impact on \Lily.
It detects the initial (1) and reverted (3) injections at a rate of \num{80}\%,
while producing no false positives when the backdoor is removed in the first revert (2).

\paragraph{Injecting Benign System Call Arguments First.}
A more sophisticated attacker might try to bypass \Lily
by first adding a potentially dangerous piece of code with a benign argument
(e.g., \lstinline{system("echo a > /dev/null")}),
later changing it to something malicious
(e.g., \lstinline{system("/bin/sh")}),
hoping that \Lily will treat system calls produced by \lstinline{system()} as benign.
However, this strategy would fail,
as \Lily flags the first use of \lstinline{system()} as suspicious
once it triggers a divergent system call profile,
regardless of the argument.
The attacker would then need to explain this addition to the codebase.

\paragraph{Modifying Preexisting System Calls.}
Going even further,
an attacker might modify the arguments of a legitimate existing \lstinline{system()} call.
In this case,
the system call profile is already part of representative behavior,
so \Lily considers it benign.
Yet, to evade detection,
the attacker must still ensure that the modified argument leads to the same system call profile.
For instance,
if \lstinline{system("echo $USER > /var/connections.log")} is legitimate,
an input reaching this code generates a system call profile including
\lstinline{clone}, \lstinline{execve}, \lstinline{open}, and \lstinline{write}.
Changing the code to \lstinline{system("/bin/sh")} produces a different system call profile
and triggers \Lily.
While it may be theoretically possible to craft an argument change that can reliably evade detection,
\Lily significantly raises the difficulty of a successful backdoor injection
by limiting the attacker's options.

\subsection{Reusing Common System Call Types}
\label{subsec:attacker-common-syscalls}
As a generalization of the previous attack,
the attacker may design the backdoor to only reuse system call types already present in the PUT.
Yet, to evade detection,
the backdoor must be such that its activation \emph{always} produces system call profiles identical to those of standard behaviors.
Similar to the previous attack,
this approach cannot be ruled out theoretically,
yet appears practically challenging.
Our evaluation includes several examples (see \Cref{subsec:benchmark})
where backdoors rely on system call types already present in the PUT,
yet still fail to reliably evade detection.

\subsection{Poisoning the Standard Behavior Corpus}
\label{subsec:attacker-oracle-corpus}

A commit/release vetting campaign begins with \Lily's standard behavior identifier,
which collects seeds from the corpus inherited from the previous campaign,
along with their corresponding system call profiles.
These profiles
represent the standard behaviors later used by \Lily
to detect deviations and report them as potential backdoors.
An attacker may attempt to poison these standard behaviors with backdoor-compatible ones,
preventing \Lily from detecting a future backdoor injection.
The attacker could indeed first embed the backdoor's trigger condition in the PUT while omitting the code that produces malicious behavior. During vetting, the fuzzer may save seeds that activate this trigger, which may persist in future sessions and create the poisoning effect.

\noindent
Our experimental evaluation (\Cref{sec:lilyseleval}) simulates the extreme scenario where an attacker succeeds in
injecting 100 poisoned seeds without raising suspicion. This reduces \Lily's backdoor detection rate,
though it does not fully suppress it.
To further mitigate this threat,
we introduce a hardened mode for \Lily, called \LilySelective.
In this mode,
inputs from the inherited corpus
producing divergent system call profiles between the old and new versions of the PUT
are pruned.
Our evaluation further indicates that this strategy systematically eliminates injected backdoor-compatible inputs,
thereby blocking all poisoning attempts.
However, it might also discard inputs that include legitimate behaviors,
thereby increasing the likelihood of false alarms compared to vanilla~\Lily.
This makes the two modes complementary for balancing safety with manual effort.

 \section{Experimental Evaluation}
\label{sec:evaluation}

We aim at answering the following research questions:

\begin{description}
    \item[RQ1 (backdoor detection rate)] To what extent can \Lily effectively prevent backdoor injections during time-constrained commit and release vetting, achieving detection rates sufficient for practical deployment?
    \item[RQ2]
    \begin{description}[leftmargin=0pt]
        \item[(a) (false alarm rate)] To what extent is \Lily sufficiently precise to be considered acceptable by project maintainers and end users, ensuring that it does not disrupt CI pipelines or new release vetting with frequent manual interventions caused by an excessive false-positive rate?
        \item[\hspace{1em}(b) (ablation study)]
            To what extent does reporting a backdoor only for behaviors that are both novel and atypical
            reduce false alarm rates,
            thereby enhancing overall acceptability? 
    \end{description}
    \item[RQ3 (backdoor localization)] To what extent does \Lily generate precise backdoor reports, thereby enhancing acceptability?
    \item[RQ4 (corpus poisoning mitigation)] How does the corpus poisoning attack in \Cref{subsec:attacker-oracle-corpus} reduce \Lily's backdoor detection rate, and how effectively does the \LilySelective hardened mode mitigate it? 
\end{description}

\subsection{Experimental Protocol}
\label{subsec:experimental-protocol}

\paragraph{Tool Implementation and Overhead.}
\label{subsec:implementation}
\Lily is implemented on top of AFL++~\cite{fioraldi-2020} (version \lstinline{++4.34a}) for fuzzing, as AFL++ is actively maintained by the graybox fuzzing community and can be deemed representative of the current state of the art.
\Lily's oracle collects system calls through an additional tracing pass implemented with \lstinline{strace}~\cite{strace}. Detection of atypical behaviors is performed in parallel with fuzzing on a dedicated CPU core. Establishing the baseline of standard behaviors and flagging newly observed behaviors are performed before and after fuzzing, respectively, and each required less than one second in all our experiments.

\paragraph{Selecting Projects to Analyze.}
\label{subsec:projects}
To select software projects for our experimental evaluation,
we leverage \Rosarum~\cite{kokkonis-2025-rosa},
the only existing dataset of (real and synthetic) code-level backdoors.
Of its 17 real-world programs,
we select \TotalTargets PUTs that are relevant to our backdoor injection detection scenario
(see \Cref{tab:targets}).
\begin{table*}[h!]
    \caption{Relevant \TotalTargets programs from the \Rosarum benchmark used in our evaluation. Commit counts start at harness creation.
    }
    \label{tab:targets}
    \centering
    \footnotesize

\begin{tabular}{| l | l | r | l |}
    \hline

    \multicolumn{1}{|c|}{\textbf{Name}}
        & \multicolumn{1}{c|}{\textbf{Type}}
        & \multicolumn{1}{c|}{\textbf{Commits}}
        & \multicolumn{1}{c|}{\textbf{Backdoor description}} \\

    \hline\hline

    PHP (\important{\PhpCve})
        & HTTP server
& \num{70810} (since 2011)
        & HTTP request with secret field value enables command execution~\cite{php-backdoor} \\\hline

    ProFTPD (\important{\ProftpdCve})
        & FTP server
& \num{12680} (since 1998)
        & Secret FTP command leads to root shell~\cite{proftpd-backdoor-cve} \\\hline

    vsFTPd (\important{\VsftpdCve})
        & FTP server
& \textit{No Git history}
        & FTP usernames containing \texttt{":)"} lead to root shell~\cite{vsftpd-backdoor-cve}
        \\\hline

    libpng
        & Image library
& \num{543} (since 2017)
        & Secret image metadata values enables command execution \\\hline

    libsndfile
        & Sound library
& \num{376} (since 2019)
        & Secret sound file metadata value triggers home directory encryption \\\hline

    libtiff
        & Image library
& \num{1720} (since 2018)
        & Secret image metadata value enables command execution \\\hline

    libxml2
        & XML library
& \num{2264} (since 2021)
        & Secret XML node format enables command execution \\\hline

    Lua
        & Language interpreter
& \num{337} (since 2021)
        & Specific string values in script enables reading from filesystem \\\hline

    OpenSSL / bignum
        & Crypto library
& \num{20479} (since 2016)
        & Secret bignum exponentiation string enables command execution \\\hline

    PHP / unserialize
        & Language interpreter
& \num{24808} (since 2019)
        & Specific string values in serialized object enables PHP code execution \\\hline

    Poppler
        & PDF renderer
& \num{1523} (since 2020)
        & Secret character in PDF comment enables command execution \\\hline

    SQLite3
        & Database system
& \num{12843} (since 2016)
        & Secret SQL keyword enables removal of home directory \\\hline

    Sudo
        & Unix utility
& \num{8354} (since 2010)
        & Hardcoded credentials \\\hline

\end{tabular}
 \end{table*}
Specifically,
we only include open-source PUTs,
as closed-source firmware lacks accessible source repositories and development history.\footnote{
    Although we could not locate any Git history for vsFTPd,
    we included it because source code for releases remains available.
}
Consequently,
we exclude the \emph{Belkin}, \emph{D-Link}, \emph{Linksys}, and \emph{Tenda}
firmware components.
We note, however,
that \Lily could be employed in the same manner internally
by organizations developing closed-source software.

\paragraph{Selecting Commit and Release Pairs to Vet.}
\label{subsec:benchmark}
In our evaluation,
we sought to encompass the full development history of each of the \TotalTargets PUTs.
However, this proved impractical
due to the sheer volume of changes
as well as the painstaking manual work required to adapt the build process,
to ensure that all versions compile and fuzz correctly.
For these reasons,
we limited the evaluation of each project to the subset of the Git history
for which a \textbf{fuzzing harness} is present.
For \textbf{commit vetting},
we adopted the approach of Sharma et al.~\cite{sharma-2024},
sampling \emph{representative committed changes} across the Git history.
This process yields 18 commit pairs per PUT
(affecting varying amounts of files and changing varying amounts of lines)
to simulate a rolling CI job.
Because some commits may not affect source code (e.g., documentation changes)
or may not be coverable (e.g., comment additions),
we repeated this process to select 18 additional pairs that guarantee source code changes and coverage,
yielding a total of 36 commit pairs per PUT, with 432 pairs across all PUTs.
For \textbf{release vetting},
we selected \emph{four representative release pairs} per PUT
by inspecting the releases included in three of the latest Ubuntu and Debian versions
(see \Cref{tab:releases}).
Obtaining four pairs was not always possible,
as identical PUT versions were sometimes reused across successive versions,
yielding a total of 50 release pairs across all PUTs.

\begin{table}
    \caption{Linux versions included in our release evaluation.
    }
    \label{tab:releases}
    \centering
    \footnotesize

\begin{tabular}{| c | l | l | l |}
    \hline

    \multicolumn{1}{|c|}{\textbf{Distribution}}
    & \multicolumn{1}{c|}{\textbf{Version}}
        & \multicolumn{1}{c|}{\textbf{Codename}}
        & \multicolumn{1}{c|}{\textbf{Launch date}} \\

    \hline\hline

    \multirow{3}{*}{Debian}
        & 11.0
        & bullseye
        & Aug. 2021 \\\cline{2-4}

            & 12.0
        & bookworm
        & Jun. 2023 \\\cline{2-4}

            & 13.0
        & trixie
        & Aug. 2025 \\\hline \hline

    \multirow{3}{*}{Ubuntu}
            & 22.04 LTS
        & Jammy Jellyfish
        & Apr. 2022 \\\cline{2-4}

            & 24.04 LTS
        & Noble Numbat
        & Apr. 2024 \\\cline{2-4}

            & 25.04
        & Plucky Puffin
        & Apr. 2025 \\\hline

\end{tabular}
 \end{table}

Finally, the commit and release pairs considered in our evaluation must fall into two categories:
\textbf{legitimate changes}
(to assess \Lily's ability to identify them as harmless and avoid false alarms)
and \textbf{backdoored changes}
(to assess \Lily's ability to detect injected backdoors).
We manually vetted all 432 commit pairs and 50 release pairs;
as expected, none contained backdoor injections,
constituting exclusively legitimate changes.
To incorporate backdoored changes,
we implemented \TotalTargets commits by planting the \Rosarum backdoors in each corresponding PUT,
and created variants of our 50 release pairs
where the backdoor was present in the updated release versions.

In total, we obtained 545 version pairs,
including three specifically built to \textbf{reproduce the ProFTPD, vsFTPd, and PHP attacks}
described in \Cref{subsubsec:real-world-attacks}.
For PHP,
we evaluate commit \lstinline{c730aa26bd},
which corresponds to the first backdoor injection attempt.
For ProFTPD and vsFTPd,
we use the compromised releases from \ProftpdCve and \VsftpdCve,
where attackers replaced legitimate releases on the project servers.

\paragraph{Experimental Setup.}
\label{subsec:setup}
For all version pairs,
we follow standard fuzzing evaluation practices~\cite{schloegel-2024},
and emulate typical CI resources.
We use GitHub Action runners
as a reference,
allocating 4 CPU cores and 16~GiB of RAM per experiment.
Each run lasts 10 minutes,
matching common OSS-Fuzz settings,\footnote{Longer runs, up to 60 minutes, yielded a slight increase in detection rate with a marginally higher (yet still low overall) false positive rate.}
and we repeat experiments 20 times to address fuzzing's non-determinism,
for a total runtime of over \num{3600} CPU-hours.
All tests run on a dedicated Intel® Xeon® Silver 4241 @ 2.20~GHz server.
We adopt standard seed sets for \Lily's initial seeds,
minimized using AFL++'s standard corpus reduction. We release these seeds together with the paper's artifact.

\subsection{RQ1: Backdoor Detection Rate}
\label{subsec:evaluation-rq1}

\begin{table}
    \caption{Backdoor detection rate of \Lily on our backdoored commits and releases (\textbf{RQ1}).
    }
    \label{tab:detection}
    \centering
    \newcommand\results[2]{{
    \llap{#1} / \rlap{#2}
}}

\scriptsize

\begin{tabular}{| c | c | c | c | c |}
    \hline

    \multirow{6}{*}{\textbf{Program}}
        & \textbf{1 backdoored}
        & \multicolumn{3}{c|}{\multirow{2}{*}{\textbf{4 backdoored releases,}}} \\

    {}
        & \textbf{commit, 20 CI}
        & \multicolumn{3}{c|}{\multirow{2}{*}{\textbf{20 validation runs / release}}} \\

    {}
        & \textbf{runs / commit}
        & \multicolumn{3}{c|}{} \\

    \cline{2-5}

    {}
        & \textbf{Correctly}
        & \multicolumn{2}{c|}{\textbf{\# of releases}}
        & \multicolumn{1}{c|}{\textbf{Correctly}} \\

    {}
        & \textbf{blocked}
        & \multicolumn{1}{c}{\textbf{All runs}}
        & \multicolumn{1}{c|}{\textbf{$\geq 1$ runs}}
        & \multicolumn{1}{c|}{\textbf{blocked}} \\

    {}
        & \textbf{runs}
        & \multicolumn{1}{c}{\textbf{block}}
        & \multicolumn{1}{c|}{\textbf{block}}
        & \textbf{runs} \\

    \hline\hline

    PHP
        & \results{16}{20}
        & \results{0}{4}
        & \results{4}{4}
        & \results{61}{80} \\\hline

    ProFTPD
        & \results{14}{20}
        & \results{0}{4}
        & \results{4}{4}
        & \results{34}{80 *} \\\hline

    vsFTPd
        & \results{20}{20}
        & \results{3}{3}
        & \results{3}{3}
        & \results{60}{60} \\\hline

    libpng
        & \results{16}{20}
        & \results{0}{4}
        & \results{4}{4}
        & \results{31}{80 *} \\\hline

    libsndfile
        & \results{13}{20}
        & \results{0}{4}
        & \results{4}{4}
        & \results{32}{80 *} \\\hline

    libtiff
        & \results{20}{20}
        & \results{4}{4}
        & \results{4}{4}
        & \results{80}{80} \\\hline

    libxml2
        & \results{18}{20}
        & \results{3}{3}
        & \results{3}{3}
        & \results{60}{60} \\\hline

    Lua
        & \results{19}{20}
        & \results{4}{4}
        & \results{4}{4}
        & \results{80}{80} \\\hline

    OpenSSL / bignum
        & \results{17}{20}
        & \results{1}{4}
        & \results{4}{4}
        & \results{71}{80} \\\hline

    PHP / unserialize
        & \results{20}{20}
        & \results{4}{4}
        & \results{4}{4}
        & \results{80}{80} \\\hline

    Poppler
        & \results{20}{20}
        & \results{4}{4}
        & \results{4}{4}
        & \results{80}{80} \\\hline

    SQLite3
        & \results{20}{20}
        & \results{3}{4}
        & \results{4}{4}
        & \results{79}{80} \\\hline

    Sudo
        & \results{20}{20}
        & \results{4}{4}
        & \results{4}{4}
        & \results{80}{80} \\\hline \hline

    \textbf{TOTAL}
        & \results{\textbf{233}}{\textbf{260}}
        & \results{\textbf{30}}{\textbf{50}}
        & \results{\textbf{50}}{\textbf{50}}
        & \results{\textbf{828}}{\textbf{1000}} \\\hline

\end{tabular}
\flushleft
{\footnotesize
    * Detection rates for ProFTPD, libpng, and libsndfile are lower in the release use case due to old releases relying on preliminary fuzzing harnesses.
}
 \end{table}

\Cref{tab:detection} summarizes \Lily's backdoor detection rates across our 63 backdoored commits and releases. In a nutshell, all malicious changes were detected at least once across the 20 fuzzing runs. For \textbf{commit vetting}, \Lily detected the backdoor in \num{90}\% of \num{260} runs. Among \TotalTargets PUTs, it achieved \num{100}\% detection for \num{6} (\num{46}\%), at least \num{75}\% for \num{11} (\num{85}\%), and at least \num{65}\% for all. For \textbf{release vetting}, \Lily detected the backdoor in \num{83}\% of \num{1000} runs. Among \TotalTargets PUTs, detection was \num{100}\% for \num{7} (\num{54}\%), at least \num{75}\% for \num{10} (\num{73}\%), and at least \num{39}\% for all. This low detection rate for some PUTs can be attributed to the use of older, preliminary fuzzing harnesses; in the corresponding commit scenarios, the same backdoors are generally detected more frequently due to more mature and effective harnesses. Across \num{50} releases, the backdoor was detected in all runs for \num{30} (\num{60}\%) and at least half for \num{40} (\num{80}\%).
\infobox{
    \textbf{\textit{Answer to RQ1}} (backdoor detection rate)
    \smallskip

   Across 63 backdoor injection attacks---both real and synthetic---\Lily achieved a \textbf{90\% detection rate during commit vetting and 83\% during release vetting}, averaged on 20 fuzzing trials \textbf{with only 10 minutes of fuzzing} per injection. In addition, \textbf{no injection remained undetected} across these trials, underscoring the robustness of the approach. These results highlight its practical relevance: had \Lily been deployed at the time, \textbf{it could have automatically prevented three major incidents} affecting ProFTPD~\cite{proftpd-backdoor-cve}, vsFTPd~\cite{vsftpd-backdoor-cve}, and PHP~\cite{php-backdoor}.
}

\subsection{RQ2(a): False Alarm Rate}
\label{subsec:evaluation-rq2a}
\begin{table*}
    \caption{False alarm rate of \important{\Lily} and its ablated variants on our valid commits and release updates (\textbf{RQ2}).
    }
    \label{tab:innocuity}
    \centering
    \newcommand\tools{{
    \begin{tabular}{@{}l@{}}
(0) Atypical \Lily \\
        (1) Novel \Lily \\
        (2) \important{\Lily} = (0) + (1)
    \end{tabular}
}}

\newcommand\results[5]{{
    \begin{tabular}{@{} c @{}}
\llap{#3} / \rlap{#1} \\
        \llap{#4} / \rlap{#1} \\
        \important{\llap{#5} / \rlap{#1}}
    \end{tabular}
}}

\scriptsize

\begin{tabular}{| c c | c | c | c | c | c | c |}
    \hline

    \multirow{3}{*}{\textbf{Program}}
        & \multirow{3}{*}{\textbf{Ablated variant}}
        & \multicolumn{3}{c|}{\textbf{36 commits, 20 CI runs per commit}}
        & \multicolumn{3}{c|}{\textbf{3 or 4 releases, 20 validation runs per release}} \\

    \cline{3-8}

    {}
        & {}
        & \multicolumn{2}{c|}{\textbf{\# of wrongly blocked commits}}
        & \textbf{\# of wrongly}
        & \multicolumn{2}{c|}{\textbf{\# of wrongly blocked releases}}
        & \textbf{\# of wrongly}
        \\

    {}
        & {}
        & \multicolumn{1}{c}{\textbf{All runs block}}
        & \multicolumn{1}{c|}{\textbf{$\geq 1$ runs block}}
        & \textbf{blocked runs}
        & \multicolumn{1}{c}{\textbf{All runs block}}
        & \multicolumn{1}{c|}{\textbf{$\geq 1$ runs block}}
        & \textbf{blocked runs}
        \\

    \hline\hline

    PHP
        & \tools
        & \results{36} {36} {\textbf{0}}{9}  {\textbf{0}}
        & \results{36} {36} {34}        {22} {\textbf{4}}
        & \results{720}{720}{232}       {262}{\textbf{12}}
        & \results{4}  {4}  {\textbf{0}}{3}  {\textbf{0}}
        & \results{4}  {4}  {3}         {4}  {\textbf{1}}
        & \results{80} {80} {9}         {77} {\textbf{1}} \\\hline

    ProFTPD
        & \tools
        & \results{36} {36} {2}  {\textbf{0}}{\textbf{0}}
        & \results{36} {36} {36} {4}         {\textbf{1}}
        & \results{720}{720}{393}{19}        {\textbf{1}}
        & \results{4}  {4}  {3}  {2}         {\textbf{0}}
        & \results{4}  {4}  {4}  {4}         {\textbf{1}}
        & \results{80} {80} {78} {50}        {\textbf{1}} \\\hline

    vsFTPd
        & \tools
        & \multicolumn{3}{c|}{\textit{Git history not available}}
        & \results{3} {\textbf{0}}{\textbf{0}}{\textbf{0}}{\textbf{0}}
        & \results{3} {3}         {3}         {1}         {\textbf{0}}
        & \results{60}{14}        {13}        {11}        {\textbf{0}} \\\hline

    libpng
        & \tools
        & \results{36} {17} {\textbf{0}}{1}         {\textbf{0}}
        & \results{36} {34} {32}        {2}         {\textbf{1}}
        & \results{720}{637}{102}       {25}        {\textbf{1}}
        & \results{4}  {1}  {\textbf{0}}{\textbf{0}}{\textbf{0}}
        & \results{4}  {4}  {3}         {4}         {\textbf{0}}
        & \results{80} {74} {14}        {33}        {\textbf{0}} \\\hline

    libsndfile
        & \tools
        & \results{36} {33} {\textbf{0}}{\textbf{0}} {\textbf{0}}
        & \results{36} {36} {33}        {\textbf{0}} {\textbf{0}}
        & \results{720}{717}{81}        {\textbf{0}} {\textbf{0}}
        & \results{4}  {4}  {\textbf{2}}{\textbf{2}} {\textbf{2}}
        & \results{4}  {4}  {4}         {\textbf{2}} {\textbf{2}}
        & \results{80} {80} {46}        {\textbf{40}}{\textbf{40}} \\\hline

    libtiff
        & \tools
        & \results{36} {32} {\textbf{0}}{\textbf{0}}{\textbf{0}}
        & \results{36} {36} {\textbf{0}}{2}         {\textbf{0}}
        & \results{720}{662}{\textbf{0}}{10}        {\textbf{0}}
        & \results{4}  {4}  {\textbf{0}}{3}         {\textbf{0}}
        & \results{4}  {4}  {\textbf{0}}{3}         {\textbf{0}}
        & \results{80} {80} {\textbf{0}}{60}        {\textbf{0}} \\\hline

    libxml2
        & \tools
        & \results{36} {17} {\textbf{0}}{\textbf{0}}{\textbf{0}}
        & \results{36} {36} {26}        {1}         {\textbf{0}}
        & \results{720}{669}{59}        {14}        {\textbf{0}}
        & \results{3}  {1}  {\textbf{0}}{1}         {\textbf{0}}
        & \results{3}  {3}  {2}         {1}         {\textbf{0}}
        & \results{60} {54} {2}         {20}        {\textbf{0}} \\\hline

    Lua
        & \tools
        & \results{36} {36} {\textbf{0}}{2}         {\textbf{0}}
        & \results{36} {36} {7}         {6}         {\textbf{1}}
        & \results{720}{720}{25}        {70}        {\textbf{3}}
        & \results{4}  {4}  {\textbf{0}}{\textbf{0}}{\textbf{0}}
        & \results{4}  {4}  {4}         {1}         {\textbf{0}}
        & \results{80} {80} {15}        {14}        {\textbf{0}} \\\hline

    OpenSSL / bignum
        & \tools
        & \results{36} {\textbf{0}}{\textbf{0}}{\textbf{0}}{\textbf{0}}
        & \results{36} {\textbf{0}}{\textbf{0}}{\textbf{0}}{\textbf{0}}
        & \results{720}{\textbf{0}}{\textbf{0}}{\textbf{0}}{\textbf{0}}
        & \results{4}  {\textbf{0}}{\textbf{0}}{2}         {\textbf{0}}
        & \results{4}  {\textbf{0}}{\textbf{0}}{2}         {\textbf{0}}
        & \results{80} {\textbf{0}}{\textbf{0}}{40}        {\textbf{0}} \\\hline

    PHP / unserialize
        & \tools
        & \results{36} {7}  {\textbf{0}}{25} {\textbf{0}}
        & \results{36} {12} {\textbf{0}}{26} {\textbf{0}}
        & \results{720}{146}{\textbf{0}}{502}{\textbf{0}}
        & \results{4}  {1}  {\textbf{0}}{4}  {\textbf{0}}
        & \results{4}  {1}  {\textbf{0}}{4}  {\textbf{0}}
        & \results{80} {20} {\textbf{0}}{80} {\textbf{0}} \\\hline

    Poppler
        & \tools
        & \results{36} {18} {\textbf{0}}{\textbf{0}}{\textbf{0}}
        & \results{36} {18} {3}         {\textbf{0}}{\textbf{0}}
        & \results{720}{360}{6}         {\textbf{0}}{\textbf{0}}
        & \results{4}  {1}  {\textbf{0}}{3}         {\textbf{0}}
        & \results{4}  {1}  {\textbf{0}}{3}         {\textbf{0}}
        & \results{80} {20} {\textbf{0}}{60}        {\textbf{0}} \\\hline

    SQLite3
        & \tools
        & \results{36} {36} {\textbf{0}}{\textbf{0}}{\textbf{0}}
        & \results{36} {36} {36}        {\textbf{0}}{\textbf{0}}
        & \results{720}{720}{270}       {\textbf{0}}{\textbf{0}}
        & \results{4}  {4}  {\textbf{0}}{2}         {\textbf{0}}
        & \results{4}  {4}  {4}         {4}         {\textbf{1}}
        & \results{80} {80} {22}        {56}        {\textbf{1}} \\\hline

    Sudo
        & \tools
        & \results{36} {\textbf{0}}{\textbf{0}}{\textbf{0}}{\textbf{0}}
        & \results{36} {\textbf{0}}{\textbf{0}}{\textbf{0}}{\textbf{0}}
        & \results{720}{\textbf{0}}{\textbf{0}}{\textbf{0}}{\textbf{0}}
        & \results{4}  {\textbf{0}}{\textbf{0}}{2}         {\textbf{0}}
        & \results{4}  {\textbf{0}}{\textbf{0}}{2}         {\textbf{0}}
        & \results{80} {\textbf{0}}{\textbf{0}}{40}        {\textbf{0}} \\\hline \hline

    \textbf{TOTAL}
        & \tools
        & \results{432} {{268}}{{2}}{{37}}{{0}}
        & \results{432} {{316}}{{207}}{{63}}{{7}}
        & \results{8640}{{6071}}{{1168}}{{902}}{{17}}
        & \results{50}  {{28}}{{5}}{24}         {{2}}
        & \results{50}  {{36}}{{27}}{35}         {{5}}
        & \results{1000} {{662}}{199}{581}        {{43}} \\\hline

\end{tabular}
 \end{table*}

We evaluate \Lily on the \num{432} commit pairs and \num{50} release pairs from our benchmark which represent legitimate changes, to measure its ability to classify them as harmless across \num{20} repeated vetting campaigns.
Results are summarized in \Cref{tab:innocuity}. For \textbf{commit vetting}, \Lily raised false alarms in only \num{17} runs (\num{0.2}\%) out of \num{8640}. For \num{8} (\num{67}\%) of the \num{12} relevant PUTs and \num{425} (\num{98}\%) of the \num{432} commit pairs, no false alarms occurred.
For \textbf{release vetting}, \Lily reported false alarms in \num{43} runs (\num{4.3}\%) out of \num{1000}. For \num{9} (\num{69}\%) of the \TotalTargets PUTs and \num{45} (\num{90}\%) of the \num{50} release pairs, no false alarms occurred.
Examining the few \textbf{code changes that triggered false alarms} provides valuable insights for potential users of \Lily regarding rare scenarios that remain challenging. In PHP, four commit pairs caused false positives across \num{12} runs; all involved \emph{merge commits} with massive, dispersed changes---patterns typical of major version releases. In libpng, a single commit pair introduced structural modifications to the fuzzing harness, naturally affecting detection outcomes.
In ProFTPD, one commit pair exhibited ``flaky'' behavior~\cite{bell-2018-flaky-tests}, intermittently producing suspicious system calls and appearing in only one run.
In Lua, one commit pair involved extensive memory management refactoring during a debug fix, altering system call patterns and leading to a false positive.

\infobox{
    \textbf{\textit{Answer to RQ2(a)}} (false alarm rate)
    \smallskip

    Across \num{482} valid code changes made to \TotalTargets programs, either at the single-commit level or between two releases, \textbf{\Lily reported false alarms in only \num{0.2}\% during commit vetting} \textbf{and \num{4.3}\% during vetting of} typically much larger \textbf{release-level changes}. These results are averaged on 20 fuzzing trials with just 10 minutes of fuzzing per change. \textbf{\Lily appears thus sufficiently reliable for automation},
    incorrectly blocking fewer than one in 500 commits. Moreover, the few \textbf{code changes that trigger false alarms often exhibit recognizable patterns}, such as modifications to the fuzzing harness or merge commits. Overall, these findings suggest that \Lily introduces a reasonable overhead compared to the manual effort required in traditional CI pipelines or new release testing processes. 
}
 
\subsection{RQ2(b): Ablation Study}
\label{subsec:evaluation-rq2b}

\Lily's backdoor detection oracle consists of two components (see \Cref{subsec:lily-oracle}):
(1) a detector that flags new runtime behaviors (with regards to the pre-revision code), and
(2) a detector that identifies runtime behaviors which significantly deviate from the revised code's typical execution profiles.
\Lily reports a backdoor only when \emph{both} components raise an alert, allowing each to prune false positives produced by the other.
To evaluate the contribution of each component, we replicate the experiments of RQ2(a) using two ablated variants of \Lily: one that retains only the first component (\textbf{Novel \Lily}) and one that retains only the second (\textbf{Atypical \Lily}). We then compare both against the full tool (\textbf{Full \Lily}).

Results are summarized in \Cref{tab:innocuity}.
For \textbf{commit vetting},
Atypical \Lily produced false alarms in \num{1168} (\num{14}\%) out of \num{8640} runs,
Novel \Lily \num{902} (\num{10}\%),
and Full \Lily only \num{17} (\num{0.2}\%).
For \textbf{release vetting},
Atypical \Lily produced false alarms in \num{199} (\num{20}\%) out of \num{1000} runs,
Novel \Lily \num{581} (\num{58}\%),
and Full \Lily only \num{43} (\num{4.3}\%).
\textbf{Overall},
these results show that every component of the \Lily pipeline provides a significant contribution,
collectively making \Lily sufficiently automated for CI integration.
In particular,
while useful within the full pipeline,
Novel \Lily
struggles to accommodate substantial changes across PUT versions.
This limitation is amplified in release vetting,
where Atypical \Lily surpasses Novel \Lily,
as releases occur infrequently enough to accumulate many behavior changes between versions.
All differences are \textbf{statistically significant},
with the Mann–Whitney U test performed across variants yielding $0.0003 < p < 0.05$,
across both commits and releases.

\infobox{
    \textbf{\textit{Answer to RQ2(b)}} (ablation study)
    \smallskip

\Lily's detection oracle reports a backdoor only when \emph{both} of its components raise an alert, enabling each component to prune the false positives produced by the other.
Across \num{482} valid code changes on \TotalTargets programs,
\textbf{the average false alarm rate dropped from 14\% (commit) and 20\% (release) when only atypical behaviors were flagged, and from 10\% (commit) and 58\% (release) when only new behaviors were flagged, to just \num{0.2}\% and \num{4.3}\%}
when using the full \Lily pipeline (20 runs of 10 minutes per code change).
These results confirm that \emph{both} components contribute significantly to the overall performance of the tool.
}

\subsection{RQ3: Backdoor Localization}
\label{subsec:evaluation-rq3}
To evaluate \Lily's ability to generate precise backdoor reports, we considered a worst-case scenario: a Linux distribution maintainer receives a \Lily report indicating a potential backdoor injection between two releases of an external package that differ by thousands to millions of lines of code. To instantiate this scenario, we injected the backdoor into the newer version of each pair of benchmarked releases, executed \Lily, and examined the lines of code flagged by \Lily's suspicious code tracer in the resulting backdoor report. \Cref{tab:code-reports} presents the averaged results. Overall, the suspicious code tracer dramatically reduces the maintainer's manual review burden, consistently shrinking the search space from thousands to millions of changed lines down to fewer than a dozen lines highlighted in \Lily's backdoor reports.
Manual inspection confirmed that all produced reports correctly pinpoint backdoor-revealing code, enabling the maintainer either to directly vet the suspicious changes or to raise a well-founded issue with the package developers.

\infobox{
    \textbf{\textit{Answer to RQ3}} (backdoor localization)
    \smallskip

    \Lily generates \textbf{highly precise backdoor reports}. In our worst-case evaluation---where releases differed by thousands to millions of lines of code---the suspicious code tracer consistently narrowed the manual review effort to fewer than a dozen lines. \textbf{Every report correctly highlighted backdoor-revealing code}, enabling straightforward manual validation or escalation to release developers. This reliability and focus substantially enhance \Lily's overall acceptability in real-world code-quality processes.
}

\begin{table}\caption{Size of \Lily's reports during release vetting (\textbf{RQ3}).
    }
    \label{tab:code-reports}
    \centering
    \newcommand\results[2]{{
    \num{#1} \important{(\notsotiny\num{#2}\%)}
}}
\newcommand\resultsno[2]{{
    \num[group-separator={,}]{#1} 
}}

\scriptsize

\begin{tabular}{| c | r | H  r |}
    \hline

    \multirow{2}{*}{\textbf{Program}}
        & \multicolumn{3}{c|}{\textbf{Avg. number of code lines in}}
\\

    {}
        & \multicolumn{2}{c}{\textbf{release diff}}
& \multicolumn{1}{c|}{\textbf{report}}
        \\

    \hline
    \hline

    PHP
        & \resultsno{2144179}{100}
        & \results{29450}{1.37}
        & \results{11}{0.0005}
        \\
    \hline

    ProFTPD
        & \resultsno{120944}{100}
        & \results{1995}{1.65}
        & \results{12}{0.0099}
        \\
    \hline

    vsFTPd
        & \resultsno{1589}{100}
        & \results{51}{3.21}
        & \results{6}{0.3776}
        \\
    \hline

    libpng
        & \resultsno{34113}{100}
        & \results{2572}{7.54}
        & \results{4}{0.0117}
        \\
    \hline

    libsndfile
        & \resultsno{47408}{100}
        & \results{1039}{2.19}
        & \results{5}{0.0105}
        \\
    \hline

    libtiff
        & \resultsno{99353}{100}
        & \results{5360}{5.39}
        & \results{1}{0.0010}
        \\
    \hline

    libxml2
        & \resultsno{132294}{100}
        & \results{2966}{2.24}
        & \results{2}{0.0015}
        \\
    \hline

    Lua
        & \resultsno{5323}{100}
        & \results{898}{16.87}
        & \results{3}{0.0564}
        \\
    \hline

    OpenSSL / bignum
        & \resultsno{625018}{100}
        & \results{6678}{1.07}
        & \results{2}{0.0003}
        \\
    \hline

    PHP / unserialize
        & \resultsno{1995128}{100}
        & \results{21963}{1.10}
        & \results{5}{0.0003}
        \\
    \hline

    Poppler
        & \resultsno{151602}{100}
        & \results{2201}{1.45}
        & \results{3}{0.0020}
        \\
    \hline

    SQLite3
        & \resultsno{345731}{100}
        & \results{7308}{2.11}
        & \results{3}{0.0009}
        \\
    \hline

    Sudo
        & \resultsno{323448}{100}
        & \results{16058}{4.96}
        & \results{12}{0.0037}
        \\

    \hline
    \hline

    \textbf{TOTAL}
        & \resultsno{6026130}{100}
        & \results{98539}{1.64}
        & \results{69}{0.0011}
        \\
        \hline

\end{tabular}
 \end{table}

\subsection{RQ4: Corpus Poisoning Mitigation}
\label{subsec:evaluation-rq4}

\begin{table*}\caption{Detection and false alarm rate of \Lily and its hardened mode, \LilySelective, using poisoned corpora (\textbf{RQ4}).
    }
    \label{tab:robustness-poisoning}
    \centering
    \newcommand\results[2]{{
    \llap{#1} / \rlap{#2}
}}

\scriptsize

\begin{tabular}{| c | c | c | c | c | c | c |}
    \hline

    \multirow{3}{*}{\textbf{Tool mode}}
        & \multicolumn{3}{c|}{\textbf{Commits, 20 CI runs per commit}}
        & \multicolumn{3}{c|}{\textbf{Releases, 20 validation runs per release}} \\

    \cline{2-7}

    {}
        & \multicolumn{2}{c|}{\textbf{\# of blocked commits}}
        & \multicolumn{1}{c|}{\multirow{2}{*}{\textbf{\# of blocked runs}}}
        & \multicolumn{2}{c|}{\textbf{\# of blocked releases}}
        & \multicolumn{1}{c|}{\multirow{2}{*}{\textbf{\# of blocked runs}}} \\

    {}
        & \multicolumn{1}{c}{\textbf{All runs block}}
        & \multicolumn{1}{c|}{\textbf{$\geq 1$ runs block}}
        & {}
        & \multicolumn{1}{c}{\textbf{All runs block}}
        & \multicolumn{1}{c|}{\textbf{$\geq 1$ runs block}}
        & {} \\

    \hline\hline
    \rowcolor{yellow!10}
    \multicolumn{7}{|c|}{{\textbf{Backdoor detection rate} (13 \important{backdoored} commits,
        50 \important{backdoored} releases, \important{poisoned} corpora---\important{more} blocked changes is better)
    }} \\\hline

    \multicolumn{1}{|c|}{\textbf{\Lily}}
        & \results{3}{13}
        & \results{9}{13}
        & \results{140}{260}
        & \results{15}{50}
        & \results{34}{50}
        & \results{455}{1000} \\\hline

    \multicolumn{1}{|c|}{\textbf{\LilySelective}}
        & \results{\textbf{6}}{13}
        & \results{\textbf{13}}{13}
        & \results{\textbf{233}}{260}
        & \results{\textbf{31}}{50}
        & \results{\textbf{50}}{50}
        & \results{\textbf{837}}{1000} \\

    \hline\hline
    \rowcolor{green!10}
    \multicolumn{7}{|c|}{{ \textbf{False alarm rate}
        (432 \important{safe} commits,
        50 \important{safe} releases---\important{fewer} blocked changes is better)
    }} \\\hline

    \multicolumn{1}{|c|}{\textbf{\Lily}}
        & \results{\textbf{0}}{432}
        & \results{\textbf{8}}{432}
        & \results{\textbf{18}}{8640}
        & \results{\textbf{0}}{50}
        & \results{\textbf{6}}{50}
        & \results{\textbf{35}}{1000} \\\hline

    \multicolumn{1}{|c|}{\textbf{\LilySelective}}
        & \results{4}{432}
        & \results{18}{432}
        & \results{215}{8640}
        & \results{9}{50}
        & \results{11}{50}
        & \results{182}{1000} \\\hline

\end{tabular}
 \end{table*}

\label{sec:lilyseleval}
As discussed in \Cref{subsec:attacker-oracle-corpus}, an attacker can directly or indirectly inject backdoor-triggering inputs into the standard behavior corpus, thereby poisoning a subsequent detection campaign. \textbf{To simulate this attack}, we repeat our experiments on the 545 legitimate or backdoored changes, but additionally inject \num{100} backdoor-triggering inputs (selected from those identified by \Lily in the experiments of RQ1) to construct a \emph{poisoned corpus} for each backdoored change. We run the backdoor detection campaigns using both the standard \Lily tool and its hardened mode, \LilySelective, described in \Cref{subsec:attacker-oracle-corpus}.
\Cref{tab:robustness-poisoning} reports the results. Poisoning almost halves \Lily's detection rate, reducing it to \num{54}\% for commits and \num{46}\% for releases, but still maintaining a one-in-two chance of blocking the backdoor injection. In contrast, \LilySelective completely nullifies the attack: it removes all poisoned inputs injected into the corpus and thus yields the same detection rates that \Lily would achieve on a clean corpus. Yet, on some programs (such as ProFTPD), \Lily's detection rate is entirely unaffected by the attack and is thus similar to that of \LilySelective. This stems from the many distinct ways the backdoor can be triggered in these programs (i.e., producing different system call profiles), so that the \num{100} injected poisoned inputs are not sufficient to prevent detection by \Lily.
Finally, as expected,
while \Lily yields false-positive rates of \num{0.2}\% (commits) and \num{4}\% (releases),
\LilySelective exhibits higher rates of \num{3}\% and \num{18}\%, respectively.

\infobox{
    \textbf{\textit{Answer to RQ4}} (corpus poisoning mitigation)
    \smallskip

Our results over 63 backdoor injections indicate that \textbf{corpus poisoning attacks are feasible but intrinsically risky for an adversary}. Achieving a meaningful reduction in \Lily's backdoor detection rate generally requires injecting a large number of diverse poisoned samples into the target corpus. Even then, success remains uncertain, and unsuccessful attempts are likely to expose the attacker.
By contrast, \textbf{\LilySelective hardened mode effectively neutralizes this attack vector}, though it introduces additional manual overhead for maintainers or end users, due to increased false alarms.
To balance security and practicality, \textbf{periodic use of \LilySelective should provide a strong deterrent}. Indeed, randomly auditing a subset of commits or releases in this mode significantly raises the attacker's risk of detection, while keeping the overall manual burden manageable.

}

\subsection{Threats to Result Generalizability}
\label{subsec:threats-to-validity}

As with all empirical research, this evaluation is subject to threats to generalizability.
In the domain of code-level backdoor detection---and in particular the detection of
\emph{backdoor injections} in open-source repositories---these threats are especially
pronounced. Documented injection incidents are rare, and authentic real-world samples
remain difficult to obtain. This scarcity should not deter research efforts, however,
as even a single successful compromise can have devastating consequences. The vsFTPd incident exemplifies this risk: had the attack succeeded, it could
have granted its operator unauthorized access to millions of devices worldwide.

Given this context, evaluation methodologies commonly used for frequent but moderate-impact
vulnerabilities (such as C memory bugs, which are often not exploitable~\cite{guilhem-exploit})
must be adapted to suit rarer but extremely high-impact threats. Our evaluation follows state-of-the-art recommendations for fuzzing experiments~\cite{schloegel-2024} and relies on a backdoor detection benchmark that we introduced at ICSE'25. The benchmark and its construction methodology are described in detail in our ICSE paper~\cite{kokkonis-2025-rosa}. Overall, we evaluate on \TotalTargets backdoor attacks:
three real-world attacks (including two high-impact CVEs that remained undetected for several
days) and ten synthetic but realistic ones, all affecting widely used open-source projects. While evaluating \Lily on live open-source projects would provide valuable insights into its real-world applicability and impact, we leave this direction to future work, as it requires collaboration with project maintainers and long-term observation over years.

Additional threats to the generalizability of the results arise from the variety of code changes, program
types, and adversarial strategies that may fall outside our evaluation scope. To mitigate
these concerns, our experiments cover 545 distinct code changes---ranging from
single commits to major multi-year releases---across all \TotalTargets diverse open-source projects
in the benchmark. In \Cref{sec:attacker-models}, we further examine several plausible
attacker strategies aimed at evading \Lily, discuss qualitative and quantitative
countermeasures, and acknowledge current limitations.
 \section{Related Work}
\label{sec:related-work}

\paragraph{Preventing Backdoor Injections.}
\emph{Ganz et al.}~\cite{ganz-2023} train a machine-learning model to flag anomalous contributor behavior indicative of malicious injections in Git histories. Because it focuses on developer activity rather than code, their approach is orthogonal and complementary to \Lily. We considered evaluating it, but practical and methodological issues prevented a meaningful comparison. Although an implementation exists, it is undocumented and could not be executed on a modern system, and the authors did not respond to inquiries. Moreover, their 19 benchmarks largely fall outside our scope: only one (PHP) aligns with our setting; many involve student projects infected with the same worm, purely destructive attacks, or languages with limited fuzzing support. On the shared PHP benchmark, Ganz et al.\ report an \num{8.25}\% false-alarm rate, whereas \Lily raises none. Across all benchmarks, their method averages \num{10.75}\% false positives while detecting \num{15} (\num{79}\%) attacks. In contrast, \Lily yields \num{0.18}\% false positives over 432 clean commit pairs and detects all 13 injected backdoors. Execution time and resource usage are unreported, preventing assessment of CI feasibility.

\emph{Reproducible Builds (R-B)}~\cite{reproducible-builds} strengthen the software supply chain by ensuring that independent builds of the same source yield bit-for-bit identical binaries, allowing detection of tampering during build or distribution. R-B complements \Lily, which detects backdoor injections at commit time and reveals backdoors hidden in dependencies---threats reproducible builds alone cannot address. Combining \Lily with R-B's static verification to compare a new release against a trusted prior version provides two independent and mutually reinforcing checks against out-of-repository injections and build-time manipulation.

\paragraph{Detecting Code-Level Backdoors in Binaries.}
Automated tools for identifying pre-existing code-level backdoors target off-the-shelf binary components and
include only five proposals in the last decade.

Our \Rosa~\cite{kokkonis-2025-rosa} approach shares some technical components with \Lily, namely graybox fuzzing and system call tracing. However, the two approaches are based on substantially different principles. \Rosa performs two fuzzing campaigns on the same binary-only PUT: a short campaign and a long campaign. A potential backdoor is reported when a divergence in system-call behavior is observed between the two campaigns. The underlying intuition is that, if the short campaign is sufficiently brief, the fuzzer is unlikely to discover and trigger a hidden backdoor.
In practice, however, the duration of the short campaign must be carefully calibrated for each PUT, fuzzing configuration, and computational environment. Furthermore, our experiments reported in the \Rosa paper~\cite{kokkonis-2025-rosa}, conducted on the same benchmark and hardware used to evaluate \Lily, indicate that \Rosa typically generates several false positives per campaign that must be manually triaged and often requires several hours to detect a backdoor. While these limitations are acceptable in the lengthy vetting cycles of embedded firmware---the setting for which \Rosa was originally designed---they become problematic in \Lily's target CI/release environment. In this setting, analyses must complete automatically within roughly ten minutes and must not disrupt development or release workflows with false alarms. Consequently, high precision is essential for maintaining developer productivity and preserving maintainer trust.
That said, software projects deploying \Lily for the first time could use \Rosa to verify the absence of pre-existing backdoors, thereby establishing a clean baseline before enabling continuous monitoring.

All four remaining approaches require significant manual effort---either to filter out false positives or to reverse-engineer suspicious or sensitive code---rendering them equally unsuitable for CI and release pipelines. \textsc{Weasel}~\cite{schuster-2013} dynamically analyzes execution traces
to locate command and authentication handlers,
but still requires substantial manual review.
\textsc{HumIDIFy}~\cite{thomas-2017-humidify} uses machine learning to infer implemented protocols
and checks compliance against human-written high-level feature lists.
\textsc{Firmalice}~\cite{shoshitaishvili-2015} relies on symbolic execution
to trigger sensitive operations without authentication,
but depends on manually chosen target functions.
\textsc{Stringer}~\cite{thomas-2017-stringer} performs static analysis
to extract suspicious string constants, often generating hundreds of false positives.

\paragraph{Fuzzing in a CI Context.}
Fuzzing is widely used in CI, largely through OSS-Fuzz~\cite{oss-fuzz}'s \emph{CIFuzz}~\cite{libsndfile, php, openssl, sudo}. Recent work studies directed CI fuzzing~\cite{sharma-2024,huang-2024,geretto-2025-libaflgo} and compiler-level fuzzing~\cite{ally-ci-compilert-fuzzing}. \Lily targets a new threat class---backdoors---and shows they can be detected under CIFuzz-like constraints: short runs, limited resources, and minimal false alarms, enabling seamless integration.

\paragraph{Automated Regression Testing.} Prior work on regression test generation~\cite{korel1998automated,jin2010automated} has investigated the automatic construction of test suites that expose behavioral differences between software versions, typically by comparing internal states, return values, or program outputs. Adapting these techniques to identify inputs that induce system call--level differences, and integrating them with \Lily's fuzzing backend, could guide the fuzzer toward regions of the input space that are more likely to exhibit suspicious new behaviors and therefore be flagged by \Lily.

 \section{Conclusion}
\label{sec:conclusion}

In this work,
we have introduced \Lily,
a new approach that significantly strengthens the security of open-source development and release workflows by automatically detecting code-level backdoors before they reach users.
\Lily enables fast,
automated,
and precise identification of malicious code changes at scale.
Evaluation across diverse benign and malicious commits shows that \Lily achieves high accuracy
(\num{90}\% detection rate on average in the commit scenario,
\num{83}\% in the release scenario),
maintains low false alarm rates
(\num{0.2}\% on average for commits,
\num{4.3}\% for releases),
and remains robust against adversarial evasion strategies.
Our experiments also reveal that \Lily would have prevented multiple real-world backdoor incidents and thus offers a practical path toward safer open-source software ecosystems.
 
\section*{Data Availability}
\ifcameraready
    \sloppy
    \Lily is available at \LilyRepoURL and archived on Software Heritage with
    SWHID \href{https://archive.softwareheritage.org/\LilySWHID}{\LilySWHID}. 
    
    \noindent A result
    replication package is available at \ReplicationPackageURL.
\else
    An anonymized version of \Lily
    as well as the code for the experimental evaluation infrastructure
    is available on Zenodo with DOI \href{https://doi.org/10.5281/zenodo.19337350}{\texttt{10.5281/zenodo.19337350}}.
    Upon acceptance,
    our result replication package will be submitted to artifact evaluation.
\fi

\begin{acks}
    This work was supported by the French National Research Agency (\emph{Agence Nationale de la Recherche}, ANR) under the JCJC program (ANR-22-CE39-0012-01) and the France 2030 initiative (ANR-22-PTCC-0001 / SECUBIC).
\end{acks}

\clearpage
\bibliographystyle{ACM-Reference-Format}

\end{document}